\documentclass[]{spie}  

\usepackage{amsmath,amsfonts,amssymb}
\usepackage{graphicx}
\usepackage[colorlinks=true, allcolors=blue]{hyperref}

\title{An on-chip astrophotonic spectrograph with a resolving power of 12,000}

\author[a]{Pradip Gatkine}
\author[a]{Nemanja Jovanovic}
\author[b]{Jeffrey Jewell}
\author[b]{J. Kent Wallace}
\author[a,b]{Dimitri Mawet}

\affil[a]{Division of Physics, Mathematics and Astronomy, California Institute of Technology, Pasadena, CA 91125, USA}
\affil[b]{Jet Propulsion Laboratory, 4800 Oak Grove Drive, Pasadena, CA 91109, USA}

\authorinfo{Further author information: (Send correspondence to P. Gatkine)\\E-mail: pgatkine@caltech.edu}

\begin{document} 
\maketitle

\begin{abstract}
With the upcoming extremely large telescopes (ELTs), the volume, mass, and cost of the associated spectrographs will scale with the telescope diameter. Astrophotonics offers a unique solution to this problem in the form of single-mode fiber-fed diffraction-limited spectrographs on a chip. These highly miniaturized chips offer great flexibility in terms of coherent manipulation of photons. Such photonic spectrographs are well-suited to disperse the light from directly imaged planets (post-coronagraph, collected using a single-mode fiber) to characterize exoplanet atmospheres.
Here we present the results from a proof-of-concept high-resolution astrophotonic spectrograph using the arrayed waveguide gratings (AWG) architecture. This chip uses the low-loss SiN platform (SiN core, SiO$_2$ cladding) with square waveguides (800 nm $\times$ 800 nm). The AWG has a measured resolving power ($\lambda/\delta\lambda$) of $\sim$ 12,000 and a free spectral range (FSR) of 2.8 nm. While the FSR is small, the chip operates over a broad band (1200 $-$ 1700 nm).  The peak on-chip throughput (excluding the coupling efficiency) is  $\sim$40\% (- 4 dB) and the overall throughput (including the coupling loss) is $\sim$ 11\% (- 9.6 dB) in the TE mode. Thanks to the high-confinement waveguide geometry, the chip is highly miniaturized with a size of only 7.4 mm $\times$ 2 mm.

This demonstration highlights the utility of SiN platform for astrophotonics, particularly,  the capability of commercial SiN foundries to fabricate ultra-small, high-resolution, high-throughput AWG spectrographs on a chip suitable for both ground- and space-based telescopes.
\end{abstract}


\section{Introduction}



Investigating the biggest questions in astronomy $-$ from uncovering  the composition and properties of exoplanet atmospheres to probing the history of large-scale structures, metal enrichment, and galaxy evolution $-$ require spectroscopy of faint targets. These measurements typically need the collecting power of some of the largest telescopes on the ground or in space. However, the diameter of the collimator of a spectrograph operating in the seeing-limited regime increases as a function of the telescope diameter ($D$). This in turn increases the size of downstream optics and a subsequent rise in the volume, mass and cost of the spectrograph proportional to $D^{2+}$.

Astrophotonics leverages photonic technologies to address these problems as well as offer novel functionalities \cite{Gatkine2019State, jovanovic2019enabling}. The  photonic  platform  of  guided  light  in  fibers  and  waveguides  has  opened  the  doors  to  next-generation  instrumentation for both ground- and space-based telescopes in optical and near/mid-IR bands. Thanks to the single-mode nature of waveguides in photonics, the photonic spectrographs operate in the diffraction-limited regime. This makes the on-chip spectrograph highly compact (chip area $\sim$ few cm$^2$) and independent of the telescope properties \cite{jovanovic2016efficiently}. The single-mode guiding of light in waveguides collapses the conventional optical setups into 2D ``optical circuits'' and allows highly flexible and unique manipulations of light (such as fine/active control of the path lengths) which are not feasible with conventional optics.  

 Various photonic spectrograph implementations have been demonstrated before, of which, the prominent architectures  include photonic echelle gratings (PEGs), arrayed waveguide gratings (AWGs), and Fourier-transform spectrometer (FTS). These are described in more detail from an astronomical perspective in a review paper\cite{gatkine2019astrophotonic}. Among these, the Arrayed Waveguide Gratings (AWG) is the most promising approach for astronomical spectroscopy from the perspective of capability and simplicity of design and fabrication \cite{gatkine2019astrophotonic}. Previous lab and on-sky demonstrations have shown the promise of AWGs to build miniaturized spectrographs for astronomical telescopes \cite{cvetojevic2012first, gatkine2017arrayed,  jovanovic2017demonstration}. The main element of an AWG is the on-chip  phased array of  single-mode waveguides  which introduce  progressive  path  differences similar to a diffraction grating.  These discrete light paths with increasing phase differences create an interference pattern at the on-chip focal curve (along a Rowland circle) with different wavelengths within a spectral order constructively interfering at different locations on the focal curve. 

Thanks  to the highly compact form factor of these chips, they offer high thermo-mechanical stability which is critical for precision / high-resolution spectrographs. The ability to control the light in individual single-mode waveguides on a chip allows flexible and innovative design strategies such as \textbf{a)} incorporating them with polarization splitters \cite{kudalippalliyalil2020low}, \textbf{b)} with nulling interferometers \cite{gatkine2019astrophotonic, norris2020first},  \textbf{c)} with frequency combs as spectroscopic calibrators \cite{obrzud2019microphotonic}, as well as \textbf{d)} with in-line OH-emission suppression filters \cite{zhu2016arbitrary, xie2018add, hu2020integrated}, to list a few. Further, these chips can be mass-produced to stack them to build multi-object spectrographs or integral-field units. 


\section{Silicon-Nitride MPW Platform}
AWGs have been implemented in various material platforms for specific applications. For astronomy, high efficiency and broadband operation are the most important factors. In the recent years, Silicon nitride (Si$_3$N$_4$) material platform (Si$_3$N$_4$ core and SiO$_2$ cladding) has emerged as a versatile platform with its broad band of transparency (400 - 2400 nm) and ultra-low propagation loss ($<$ 0.1 dB/cm) \cite{bland2006instruments, blumenthal2018silicon}. Further, the high-index contrast ($n_{core}$ = 2.0, $n_{cladding}$ = 1.45) means low bending losses in waveguides with small radii of curvature, thus allowing an ultra-compact design.  

Commercial foundries offer a reliable, repeatable, and high-quality fabrication, thanks to the multi-decade legacy of silicon photonics. With the growth of silicon nitride platform and the relevant fabrication techniques over the last decade, commercial foundries (eg: Ligentec, Lionix, LETI, AMF, etc.) have started to offer  fabrication runs for rapid SiN prototyping. These runs, called multi-project wafer runs (MPW), offer standard thickness of silicon nitride layer and collate multiple projects on a single wafer to reduce the cost. There are two main classes of SiN thickness currently offered by SiN MPW foundries, 800 nm thick (better suited for square waveguides) and 100-200 nm thick (for rectangular waveguides). In our previous work, we simulated the high-resolution (R $\sim$ 10,000) AWGs in both of these MPW classes to examine the impact of phase errors on the AWG performance and thereby, provided fabrication tolerances to achieve high-efficiency and high spectral resolution \cite{gatkine2021potential}. 

In this paper, we present experimental results of a proof-of-concept AWG to test the suitability of the Ligentec 800-nm SiN platform for building high-resolution (R $>$ 10,000) astronomical spectrographs. The key aspects that we measured here are on-chip and coupling efficiency, spectral resolution, and broadband operation. We also examine whether the typical phase errors in this platform are small enough to ensure the desired spectral performance.

\section{Design and simulations}

\begin{figure} [ht]
   \begin{center}
   \begin{tabular}{c} 
   \includegraphics[height=5cm]{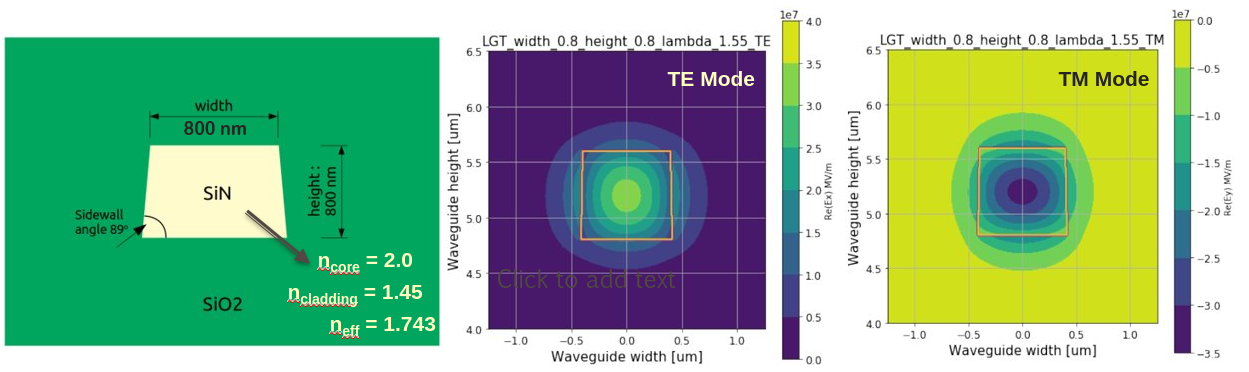}
   \end{tabular}
   \end{center}
   \caption[Full setup of an integrated photonic spectrograph] 
   { \label{fig:WG_prop}
   \underline{\textbf{Left}}: The geometry of the waveguide fabricated using the Ligentec MPW platform is shown here. The waveguide cross-section is 800 nm $\times$ 800 nm, with a sidewall angle of 89$^\circ$. 
   The core  (SiN) refractive index is 2.0, cladding (SiO$_2$)  index is 1.45, and the effective index of the waveguide ($n_{eff}$) is 1.743. 
   \underline{\textbf{Center and Right}}: The mode profile of the fundamental TE and TM (transverse electric and transverse magnetic) modes of the waveguide. The boundary of the waveguide core (SiN) is shown in orange, whereas the gradient shows the electric field mode profile at $\lambda$ = 1550 nm. The mode diameter is 1.54 $\times$ 1.42 $\mu m$ for the TE polarization and 1.50 $\times$ 1.54 $\mu m$ for the TM polarization. The numerical aperture of the waveguide is 1.34 for both TE and TM.}
   \end{figure}

\paragraph*{\underline{Waveguide geometry:}} Since this AWG was designed for the Ligentec 800 nm platform (AN800), a square-like waveguide geometry was used as shown in Fig. \ref{fig:WG_prop}. The AN here stands for the all-nitride LGT technology
platform. The core of the waveguide is a trapezoid made of SiN with a thickness of 800 nm and a  side-wall angle of 89 degrees.
The surrounding cladding is made of silicon
dioxide (SiO$_2$). The mode-profile of the waveguide for TE and TM polarizations (transverse electric and transverse magnetic) are shown in Fig. \ref{fig:WG_prop}. 
The AN800 process of Ligentec is optimized for operation around 1550 nm (the C-band in the telecom industry) with an estimated propagation loss of $<$0.2 dB/cm.  However, the waveguide operates in the single-mode regime and is transparent over the 1200-1700 nm waveband. The mode diameter is 1.54 $\times$ 1.42 $\mu m$ for the TE polarization and 1.50 $\times$ 1.54 $\mu m$ for the TM polarization. The numerical aperture of the waveguide is 1.34 for both TE and TM.

\begin{figure} [ht]
   \begin{center}
   \begin{tabular}{c} 
   \includegraphics[height=10cm]{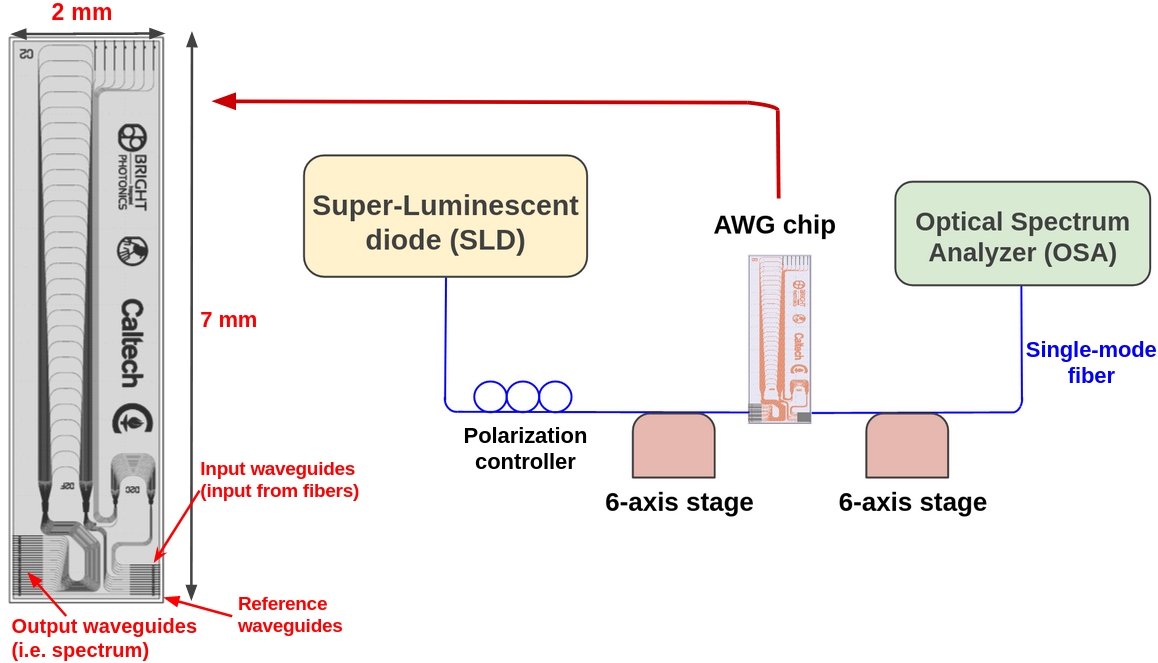}
   \end{tabular}
   \end{center}
   \caption[Full setup of an integrated photonic spectrograph] 
   {\label{fig:Characterization_scheme} The schematic of the AWG characterization setup is shown here (on the right). The light from the super-luminescent diode (SLD) is carried to the AWG chip using a polarization controller. An in-line polarization controller is used to tune the polarization. The 6-axis stages (translation $+$ rotation) allow precise alignment ($<$ 0.1 $\mu m$ tolerance) of the fiber with input/output waveguides. The light from the AWG is measured using the optical spectrum analyzer (OSA).  
}
   \end{figure}

\begin{figure} [ht]
   \begin{center}
   \begin{tabular}{c} 
   \includegraphics[height=8cm]{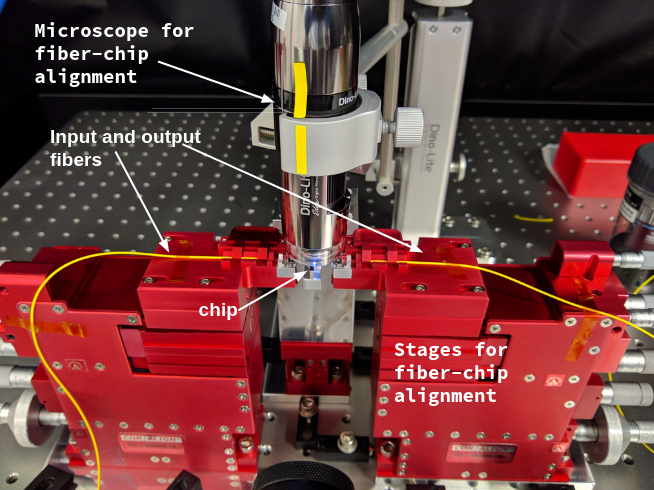}
   \end{tabular}
   \end{center}
   \caption[Full setup of an integrated photonic spectrograph] 
   { \label{fig:Characterization_pic} 
   The characterization assembly is shown here. The red colored stages are used for aligning the fibers with the waveguides. The 900x magnification microscope is used to zoom in on the waveguide-fiber interface for alignment. 
}
   \end{figure}

\begin{figure} [ht]
   \begin{center}
   \begin{tabular}{c} 
   \includegraphics[height=4.5cm]{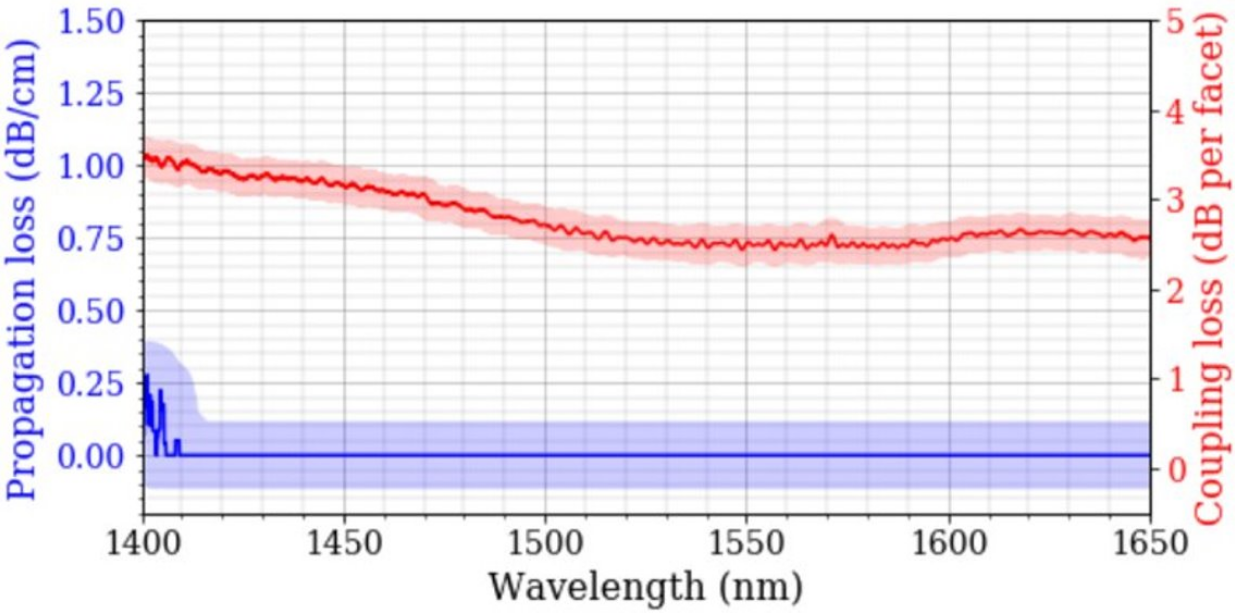}
   \end{tabular}
   \end{center}
   \caption[Full setup of an integrated photonic spectrograph] 
   { \label{fig:LGT_wg_loss} The coupling and propagation losses as a function of wavelength are shown in red and blue colors, respectively. The propagation loss for the Ligentec waveguides is $\leq$ 0.1 dB/cm in the 1450-1650 nm band. The coupling losses are measured between the on-chip inverted taper (mode diameter $\sim$ 2.5 $\mu m$) and the SMF28 fiber (mode diameter $\sim$ 10 $\mu m$). Due to their significant mode-size mismatch, the coupling losses are $\sim$ 2.5 dB/facet in the 1550 - 1650 nm band. 
} 
   \end{figure}

It can be seen that mode size of the waveguides (1.5 $\mu m$) is not suitable for coupling to the typical single-mode fibers (10 $\mu m$). To increase the mode diameter, the waveguide width is progressively reduced in an adiabatic taper (called inverted tapers), which leads to a weakly confined, and therefore a large mode size, which is better-matched to the fiber\cite{zhu2016ultrabroadband, gatkine2016development}. The Ligentec inverted tapers have a mode diameter of 2.5 $\mu$m. While these are not perfectly suited for the SMF28 fiber, they offer a better-matched solution compared to the actual waveguides. Further, ultra-high numerical aperture fibers or lensed fibers with 2.5$-$3 $\mu m$ mode diameter can be used to achieve high coupling efficiency with these tapers. The propagation loss of the waveguides is very low ($\sim$ 0.1 dB/cm) and the coupling loss with SMF28 fiber is $\sim$ 2.5 dB/facet. These were measured with multiple reference waveguides of different lengths using our characterization setup (shown in Fig. \ref{fig:Characterization_scheme} and \ref{fig:Characterization_pic}). The propagation and coupling losses (with SMF28 fiber) are shown in Fig. \ref{fig:LGT_wg_loss}.  

The material and mode profile and the design. 
Design of the AWG 
(waveguide dimesnions etc)
(No. of inputs and outputs)
(Dimensions of the waveguide)
The footprint.

\paragraph*{\underline{Specifications:}} This proof-of-concept AWG was designed to operate in the near-IR H-band (1450 - 1650 nm) and optimized for operation in TE polarization. The output channel spacing (for the discrete output waveguides) was designed to be 0.12 nm at the central wavelength, with a free spectral range (FSR) of 2.88 nm. Thus, the total number of output waveguides were designed to be 24 ($=$ 2.88 / 0.12). Note that each spectral order (i.e. 1 FSR) would have 24 spectral channels (Output waveguide 1 to output 24) and the next spectral order would again start from output waveguide 1. Thus, in an AWG,  the spectral orders overlap on each other and need to be separated using cross-dispersion setups in the future to implement a fully integrated spectrograph\cite{gatkine2020development}.

\paragraph*{\underline{AWG parameters:}} The AWG was designed using the method prescribed in the seminal paper by Meint Smit\cite{smit1996phasar}. This AWG has 24 output waveguides as mentioned before and 15 input waveguides to allow a full characterization of the off-centered inputs (beyond the scope of this paper). The AWG CAD layout is shown in Fig. \ref{fig:Characterization_scheme}. A total of 28 waveguides were used in the array. The device was fabricated at Ligentec using their AN800 process, which uses stepper lithography for fabrication. 
The gap between the waveguides at the free propagation region (FPR) is 300 nm, which is the minimum allowed distance between two SiN features on the LGT MPW platform, given the limitations of the lithography. Tapers were used at the FPR-waveguide interfaces to allow a smooth transition of refractive index from the FPR slab to the narrower waveguides, and thus minimize the reflection loss at the interface. \textbf{This device has an ultra-compact footprint of only 2 mm $\times$ 7 mm.} 

\begin{figure} [ht]
   \begin{center}
   \begin{tabular}{c} 
   \includegraphics[height=15.0cm]{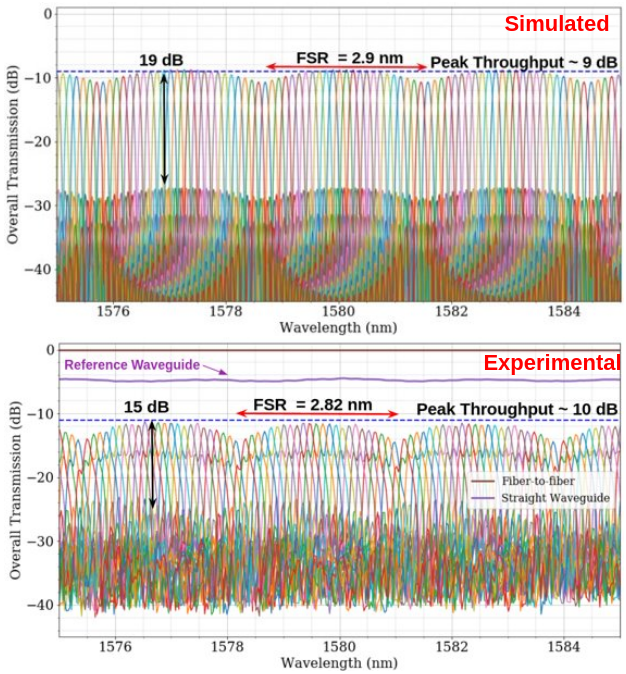}
   \end{tabular}
   \end{center}
   \caption[Full setup of an integrated photonic spectrograph] 
   { \label{fig:AWG_Transmission} 
\underline{Top panel: } The simulated AWG transmission profile in the 1575 - 1585 nm range is shown here. Three spectral orders span this range, each with an FSR = 2.9 nm. Each color shows the power in an AWG output waveguide (i.e. channel) as a function of wavelength. The simulated transmission includes a 2.5 dB/facet coupling loss. The simulated peak throughput in this band is -9 dB (12.6 \%) and the crosstalk is 19 dB.  \underline{Bottom panel: } The experimental AWG transmission profile (normalized to the input spectrum) is shown here in the same waveband. The observed FSR of the AWG is 2.82 nm, with a typical crosstalk of 15 dB. The peak throughput in this range is $\sim$ -10.5 dB (9\%) in this waveband, which includes the on-chip losses and coupling losses. The reference waveguide transmission is also shown in purple. The input spectrum (fiber-to-fiber) for this characterization is shown in brown.   
}
   \end{figure}

\paragraph*{\underline{Simulations:}} The AWG design was simulated using electric field solvers to calculate the expected AWG transmission profile as a function of wavelength. This profile is shown in the top panel of Fig. \ref{fig:AWG_Transmission}. We only simulate the on-chip insertion losses that originate due to the AWG design and from the losses at the interfaces between the FPRs and the waveguide array.
The insertion loss depends on the gap between the waveguides at the waveguide-FPR interface and the index contrast of the waveguides. Lower the index contrast, weaker the mode confinement, and thus higher the amount of light that gets coupled from the FPR slab to the waveguides. After simulation, we add 2.5 dB/facet coupling loss between the SMF28 fiber and the waveguide inverted taper to calculate the overall throughput. The simulation results in the 1575-1585 waveband are shown in the top panel of  Fig \ref{fig:AWG_Transmission}. The simulated on-chip throughput was $\sim$ -4 dB (= 40\%), and after including the coupling losses, the overall simulated throughput is $\sim$ -9 dB (= 12.5\%). The inter-channel crosstalk was observed to be $\sim$ 19 dB ($\sim$ 1.2\%), and therefore, each channel is expected to be well-isolated from the neighboring channels.  

\section{Characterization}\label{sec:characterization}
The AWG characterization setup is shown in Fig. \ref{fig:Characterization_scheme}. A super-luminescent diode (SLD) with a continuum emission over 1400 $-$ 1650 nm was used as the input source to the AWG. The polarization of the light entering the AWG was controlled using an in-line polarization controller (Thorlabs, FPC561). The input and output fibers were butt-coupled to the input and output waveguides on the chip using the precision 6-axis stages (XYZ and yaw-pitch-roll control). The broadband light source was measured to be steady as a function of time within 0.05 dB. The input fiber was coupled to the central input waveguide of the AWG. The output fiber was coupled to each of the AWG output channels one-by-one and the overall transmission response of each channel was measured using the Optical Spectrum Analyzer (OSA: Thorlabs OSA202C). A picture of the 6-axis stages with the fibers, chip, and the microscope (to aid the alignment) is shown in Fig. \ref{fig:Characterization_pic}.

\section{Experimental results}
\paragraph*{\underline{Throughput}} The overall throughput of the AWG was measured using the characterization setup described in Section \ref{sec:characterization}. The overall throughput as a function of wavelength in the 1575 - 1585 band is shown in the bottom panel of Fig. \ref{fig:AWG_Transmission}. The throughput is measured to be $-10.5$ dB in this range, which is only 1.5 dB in excess of the simulated transmission. We also see that the observed crosstalk is $\sim$ 15 dB = 3.1\%, as opposed to the simulated 19 dB. This still indicates that the AWG spectral channels are well-separated without significant inter-channel interference. The excess loss and the excess crosstalk are primarily due to the phase errors that are inevitable due to fluctuations in the fabrication process. An increase in the phase errors reduce the AWG throughput and increase the inter-channel crosstalk (which was studied in detail in our recent paper) \cite{gatkine2021potential}. Given that the additional degradation in throughput is only 1.5 dB and the crosstalk is also within acceptable limits, we can conclude that the phase errors are within limits ($leq$ 120 deg\cite{gatkine2021potential}), and hence do not degrade the AWG performance significantly. This indicates that the Ligentec platform is suitable and sufficiently stable for fabricating more realistic high-resolution AWGs (with larger FSRs) in the future.

\begin{figure} [ht]
   \begin{center}
   \begin{tabular}{c} 
   \includegraphics[height=4.75cm]{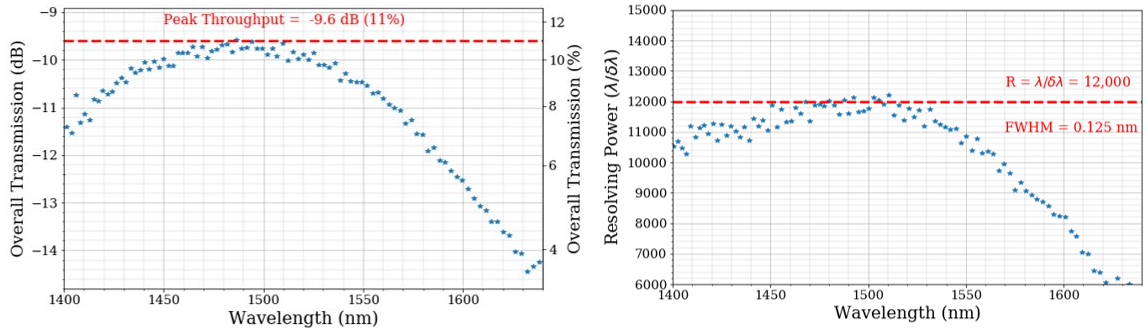}
   \end{tabular}
   \end{center}
   \caption[Full setup of an integrated photonic spectrograph] 
   { \label{fig:AWG_prop} 
   \underline{Left:} The overall transmission efficiency of the AWG as a function of wavelength is shown here in dB on the left axis, and in percentages on right axis. This includes the on-chip losses as well as the fiber-waveguide coupling losses (using SMF28 fiber). A peak throughput of 11\% (-9.6 dB) is  achieved in the range of 1450 - 1550 nm band. Note that the coupling losses account for 5.5 dB in the transmission and will be significantly reduced with the use of high-numerical-aperture fibers.   \underline{Right:} The resolving power (R = $\lambda/\delta\lambda$) of the AWG as a function of wavelength. The $\delta\lambda$ is taken as the FWHM of the discrete AWG spectral channel. The resolving power remains $>$ 10,000 in the 1400-1560 nm band. 
}
   \end{figure}

Further, we plot the overall transmission efficiency of the AWG as a function of wavelength in Fig. \ref{fig:AWG_prop}. The peak throughput is 11\% (-9.6 dB)  and a throughput of $>$ 8\% (-11 dB) is  achieved over the 1400 - 1550 nm band. Note that the coupling losses account for 5.5 dB in the transmission and will be significantly reduced with the use of high-numerical-aperture fibers or lensed fibers.  

\paragraph*{\underline{Resolving power}}
As shown in the bottom panel of Fig. \ref{fig:AWG_Transmission}, the FSR of the AWG was observed to be 2.82 nm. As such the observed channel spacing is therefore 0.1175 nm, which is close to the design spacing of 0.12 nm. We also plot the resolving power of the AWG over the 1400 $-$ 1650 nm band in Fig. \ref{fig:AWG_prop}. The resolving power (R) here is calculated as the $\lambda/\delta\lambda$, where $\delta\lambda$ is the full-width at half-maximum (FWHM) of the AWG spectral profile. We observe that the resolving power is $>$ 10,000 over a broad band of 1400 $-$ 1570 nm. The resolving power decreases at longer wavelengths due to the larger mode profiles at longer wavelengths, which result in broader spectral profiles of the channels. The high-resolving power obtained over a broad band further corroborates that the Ligentec fabrication processes are stable enough to keep the phase errors within limits \cite{gatkine2021potential}, and thus ensure the spectral properties close to the design properties.

\paragraph*{\underline{Broadband Operation}}
In this paper, we have only reported the throughput and resolving power results in the 1400 $-$ 1650 nm range here. This is restricted primarily due to the waveband of the input SLD source. However, this AWG can, in fact, operate over 1250-1700 nm due to the single-mode nature of the waveguides over this broad band. We measured the AWG transmission peaks in the  1250-1700 nm range using a supercontinuum fiber laser source (SuperK Extreme). Since the pulse repeatation rate of the SuperK source is not campatible with the Thorlabs OSA (which is a Fourier Transform Spectrograph), we cannot reliably measure the magnitude of the source spectrum. Therefore, we only show the raw output from the AWG. Regardless, it can be clearly seen that this AWG works as a `wavelength disperser' over a broad band of 1200 $-$ 1700 nm, thanks to the single-mode nature of the waveguide over this broad band, further highlighting the utility of SiN material for constructing broadband AWG spectrographs.

   \begin{figure} [ht]
   \begin{center}
   \begin{tabular}{c} 
   \includegraphics[height=5.5cm]{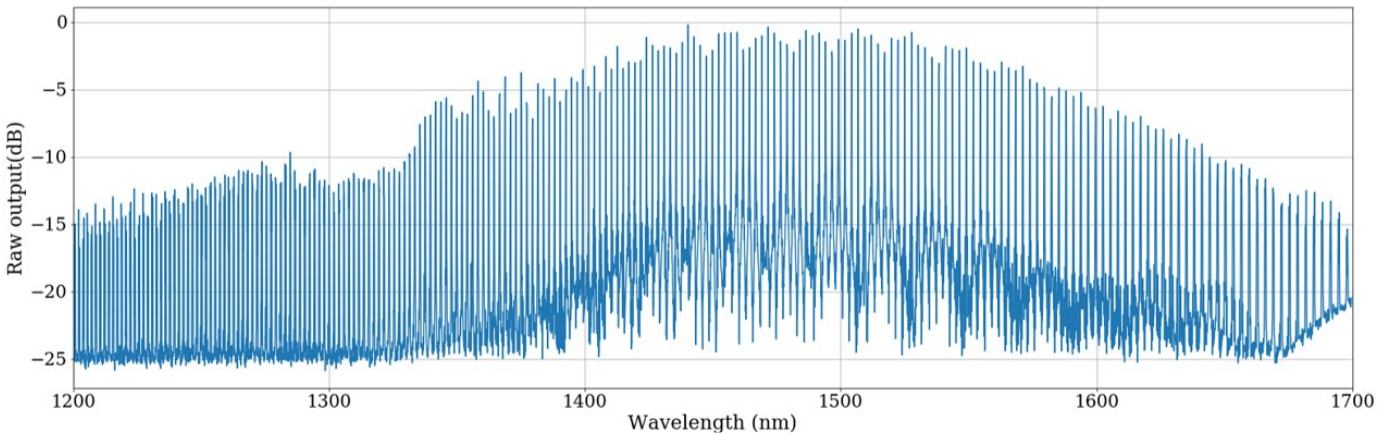}
   \end{tabular}
   \end{center}
   \caption[Full setup of an integrated photonic spectrograph] 
   { \label{fig:Broadband_transmission} 
The broadband spectral dispersion achieved by the AWG chip is shown here. The central output channel of the AWG is shown as an example. The spectral filtering action of the AWG can be seen from 1200 to 1700 nm, thus confirming the broadband nature of the SiN AWG presented here. The raw transmission output of the AWG is shown here (and not normalized to the input) due to difficulty in the measurement of input continuum of the SuperK source with the Thorlabs OSA.
}
   \end{figure}

\section{Conclusion and Future work}
In this paper, we present the results from a proof-of-concept high-resolution AWG (R $\sim$ 12,000) to show that Si$_3$N$_4$ is a promising material platform for future development of high-resolution astrophotonic spectrographs. Further, the close match of the simulated and the experimental results show that the commercial MPW fabrications are stable enough to keep the phase errors low. Thus, they are suitable for the production of high-resolution AWGs which are highly sensitive to the degradation due to phase errors. 

We further observe that the propagation loss of these waveguides is low ($\sim$ 0.1 dB/cm). The overall AWG throughput was found to be $\sim$ 11\% at 1500 nm. The major contributor to the overall loss is the coupling loss to the SMF28 fiber ($\sim$ 2.5 dB/facet). However, this can be overcome by using an ultra-high numerical aperture (UHNA) fiber or a lensed fiber with a matched numerical aperture. Thus, the SiN platform looks promising for building high-throughput astrophotonic spectrographs. 

The ultra-small footprint of the AWG chip (7.4 mm $\times$ 2 mm) shows the high level of miniaturization that can be achieved using photonics, and particularly, the high-contrast SiN platform. Here, we also show that this AWG has a broadband operation (1200 $-$ 1700 nm). This will be directly useful in the future construction of a broadband astrophotonic AWG spectrograph over the astronomical J- and H-bands (1100 $-$ 1700 nm). Thus, we corroborate the suitability of the SiN MPW platform for high-resolution astrophotonic AWGs in terms of throughput, phase errors, miniaturization, and broadband operation.

\paragraph{\underline{Future work:}}
While we demonstrate a high resolving power, this preliminary AWG has an FSR of only 2.82 nm. The next step is to build an AWG with a significantly higher FSR, integrate it with a cross-dispersion setup (for separating the spectral orders), and thus demonstrate a full-fledged deployable spectrograph. The key challenge for the upcoming work is realizing a high efficiency, integrated AWG spectrograph. The on-chip fabrication tolerances required to construct such a spectrograph, and the necessary steps to improve the fiber-chip coupling efficiency and on-chip efficiency, are discussed in detail in Gatkine et al. 2021 \cite{gatkine2021potential}. These steps primarily involve using a thinner SiN layer and inverted tapers for optimized mode matching with single-mode fibers. Such a mode-matching will reduce the propagation losses to $<$ 0.5 dB/facet and, thus, help achieve a total throughput of $>$50\%. This will be an important milestone towards leveraging the  flexibility and miniaturization offered by the AWG for on-sky deployments such as a single / multi-object spectrographs, integral field units, or an interferometer disperser for ground- and space-based telescopes.

\acknowledgments 
Pradip Gatkine was supported by the David and Ellen Lee Postdoctoral Fellowship at the California Institute of Technology. Nemanja Jovanovic acknowledges the help and expertise of BRIGHT Photonics and Ligentec in the design and fabrication of one of the chips tested in this work as part of their MPW service. This work was supported by the Wilf Family Discovery Fund in Space and Planetary Science, funded by the Wilf Family Foundation. This research was carried out at the California Institute of Technology and the Jet Propulsion Laboratory under a contract with the National Aeronautics and Space Administration (NASA) and funded through the President’s and Director’s Research $\&$ Development Fund Program.
\bibliography{report} 
\bibliographystyle{spiebib} 

\end{document}